# Improvement in Retention Time of Capacitorless DRAM with Access Transistor

Md. Hasan Raza Ansari, *Student Fellow, IEEE*, Jawar Singh *Member, Senior IEEE*

*Abstract*—In this paper, we propose a Junctionless (JL)/Accumulation Mode (AM) transistor with an access transistor (JL in series with JL/AM transistor) based capacitorless Dynamic Random Access Memory (1T-DRAM) cell. The JL transistor overcomes the problem of ultrasharp *pn* junction associated with conventional Metal-Oxide-Semiconductor (MOS) in nanoscale regime. The access transistor (AT) is utilized to reduces the leakage, and thus, improves the Retention Time (*RT*) and Sense Margin (*SM*) of the proposed capacitorless DRAM cell. Thus, the proposed DRAM cell achieved a maximum *SM* of ~4.6 µA/µm with *RT* of ~6.5 s for a gate length ($L_g$) of 100 nm. Further, this topology shows better gate length scalability with fixed gate length of AT and achieves *RT* of ~100 ms and ~10 ms for a scaled gate length of 10 nm at 27 °C and 85 °C, respectively.

*Index Terms*—Junctionless, Accumulation Mode, Temperature, 1T-DRAM, BTBT, Access Transistor.

## I. INTRODUCTION

INCREASE in demand of fast, dense, and robust memory for applications ranging from Artificial Intelligence (AI) to Internet of things (IoT) have fueled the improvement in Metal Oxide Semiconductor (MOS) devices [1]–[4]. However, the major concerns are scaling of the transistors in nanoscale regime, which degrades the performance of memory cell due to short-channel-effects and band-to-band-tunneling, and also, the formation of ultrasharp *pn* junctions. These issues can be resolved with the junctionless transistors (JLTs), which consist of same type of dopant throughout the semiconductor film [5]. The recent developments in conventional one-transistor one-capacitor (1T-1C) DRAMs show improved performance and low power consumption, however, poor scaling of capacitor limits the cell density [6].

Recently, JLTs are exploited for capacitorless DRAMs with an enhanced performance [7]–[10]. However, the Retention Time (*RT*) of these JLT based 1T-DRAMs is lower than 64 ms at 85 °C (specified by the ITRS [6]). Thus, in this paper, we have proposed a JLT based 1T-DRAM with Access Transistor (AT) that exhibits significantly (~100 x) improved *RT* against the ITRS prediction, and better scalability for higher cell density. Further, use of AT suppresses the leakage current, thus, enhances the retention characteristics of the proposed memory cell in contrast to conventional 1T-DRAM [11], [12].

Manuscript received XXX. This work was supported by Department of Science and Technology (SERB), Government of India. Grant # EMR/2016/004065.

Md. Hasan Raza Ansari and Jawar Singh are with the Electrical Engineering Department, Indian Institute of Technology Patna, Bihta, 801103, India, (e-mail: jawar@iitp.ac.in).

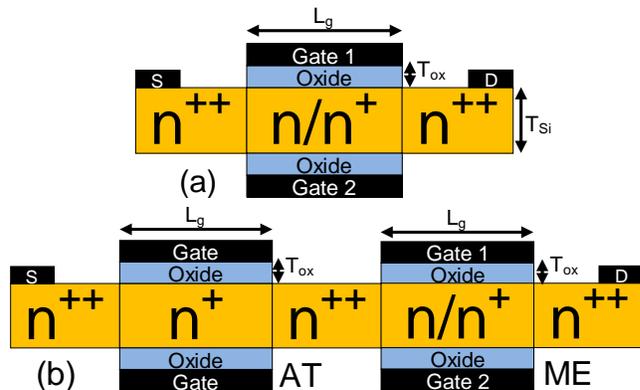

Fig. 1. Schematic diagram of 1T-DRAM (a) conventional double gate Junctionless (JL)/Accumulation Mode (AM) transistor based, and (b) proposed capacitorless DRAM (memory element) with AT.

## II. PROPOSED CAPACITORLESS DRAM

Fig. 1(a) shows the conventional double gate AM (*n*) and JL ($n^+$) devices, which are doped with same type of dopants throughout the semiconductor, with channel doping ($N_d$) of $10^{19}$ cm$^{-3}$, source and drain doping of $10^{20}$ cm$^{-3}$, gate length ($L_g$) of 100 nm along with an underlap length of 10 nm, film thickness ($T_{Si}$) of 10 nm, top and bottom oxide ($T_{ox}$) thickness of 1 nm, and gate workfunction ($\varphi_m$) of 5.0 eV (to deplete the carriers from silicon film for a positive threshold voltage of the device). The proposed capacitorless DRAM is shown in Fig. 1(b) having JL/AM transistor as a memory element (ME) in series with access transistor (AT). The AT is utilized to improve the retention characteristic and sense margin of the proposed capacitorless DRAM. The device parameters for AT are $L_g$ of 100 nm, $T_{Si}$ of 10 nm with channel doping of and $10^{19}$ cm$^{-3}$ and source/drain of $10^{20}$ cm$^{-3}$. The device parameters of ME are $n^{++}$ source/drain doping of $10^{20}$ cm$^{-3}$, $L_g$ of 100 nm and varied to 10 nm, $T_{ox}$ of 1 nm with SiO$_2$ as the dielectric layer, and width ($W_{Si}$) as 1 µm, $T_{Si}$ of 10 nm. The capacitorless DRAM operates with independent gate operations, where the front gate (Gate 1) is utilized to regulate the conduction during read, while the back gate (Gate 2) is associated with charge storage and retention [13]–[15].

The proposed capacitorless DRAM was simulated with Silvaco ATLAS simulation tool [16], with physical models of Band-To-Band-Tunneling (BTBT) and SELB impact ionization model along with doping and temperature dependent SRH models, Lombardi mobility model and bandgap narrowing model, which are well calibrated with the experimental data [17], as shown in Fig. 2(a). Fig. 2(b) demonstrates the memory characteristic with variation of read current (state "1" and "0") as a function of hold time. It is evident that incorporation of an AT with AM/JL transistor is advantageous for the proposed capacitorless DRAM cell in terms of Sense Margin (*SM*) and retention time. The AT controls generation and recombination of holes, which keeps the holes for longer duration in potential well and enhances the *SM* as well as *RT* of the proposed DRAM cell. The difference between state "1" and "0" current is termed as *SM* when it reaches to 50 % of its maximum value, and the time elapsed is referred as *RT*. Simulations results show improvement in *SM* as well as in *RT* of the memory with same device parameters and bias scheme. The incorporation of access transistor increases the *SM* and *RT* of by a factor of ~4.3 compared to conventional double gate AM.

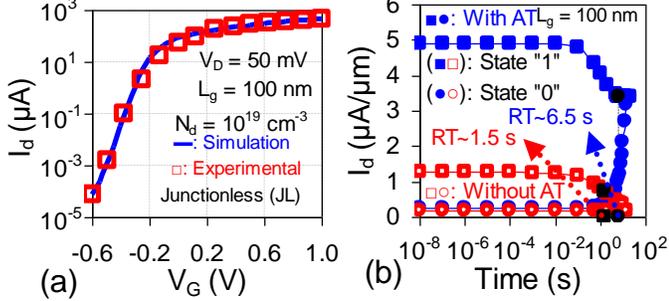

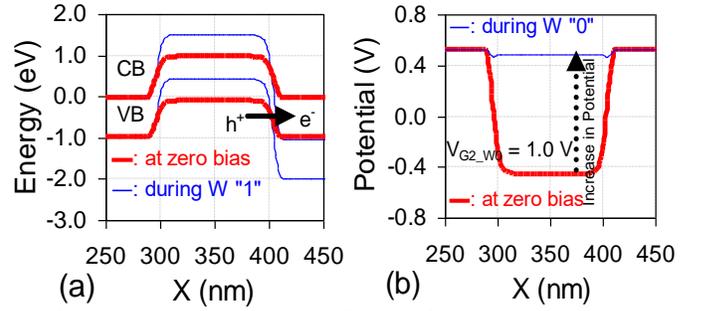

Fig. 3. (a) Variation in energy band diagram of the memory element during zero bias and Write "1" operation with gate length of 100 nm. (b) Variation in potential profile during zero bias and Write "0" with write zero bias ($V_{G2\_W0}$) of 1.0V. CB and VB indicate conduction band and Valance Band energy. $h^+$ stand for holes while $e^-$ for electrons.

Fig. 2. (a) Comparison of simulated drain current ($I_d$) - gate voltage ($V_G$) characteristics of a JL MOSFET with experimental data [17]. (b) Variation in state "1" and "0" current with hold time to estimate the retention time of the proposed memory cell with/without AT and same device parameters.

## III.  RESULTS AND DISCUSSION

The proposed capacitorless DRAM operation is based on the holes distribution in potential well (energy barrier) of the memory element (ME) JLT. The biasing scheme and timing details for different operations are illustrated in Table I. Fig. 3(a) shows the variation of energy band diagram of the ME during zero bias and write "1" operation. Energy band diagram at zero bias represents the potential well or charge storage of the ME JLT. The cutlines are taken at 1 nm above of the back gate oxide. Write "1" operation is performed with BTBT phenomenon by applying a negative bias of -1.6 V at the back gate (G2) and 1.5 V at the drain of the ME JLT. The applied bias reduces the tunneling width between back gate and drain region, which allow electrons to tunnel from valence band to conduction band from gate region to the drain region (Fig. 3(a)) and generate the electron-hole pairs. Write "0" operation is performed with forward bias mechanism by applying a positive bias at the back gate of ME JLT, which allows the stored holes to recombine with source and drain electrons (Fig. 3(b)). During write and read operations, the AT is turned on with a gate bias of 1 V and source terminal is grounded.

TABLE I
BIASING AND TIMING DETAILS OF THE PROPOSED DRAM

| Operation | $V_{G1}$ (V) | $V_{G2}$ (V) | $V_G^*$ (V) | $V_S$ (V) | $V_D$ (V) | Time (ns) |
|---|---|---|---|---|---|---|
| Write "1" | 1.0 | -1.6 | 1.0 | 0.0 | 1.5 | 10 |
| Hold | 0.0 | -0.1 | 0.0 | 0.0 | 0.0 | --- |
| Read | 1.2 | 0.1 | 1.0 | 0.0 | 0.1 | 10 |
| Write "0" | 1.0 | 1.0 | 1.0 | 0.0 | 0.0 | 10 |

* $V_G^*$ is used for proposed architecture.

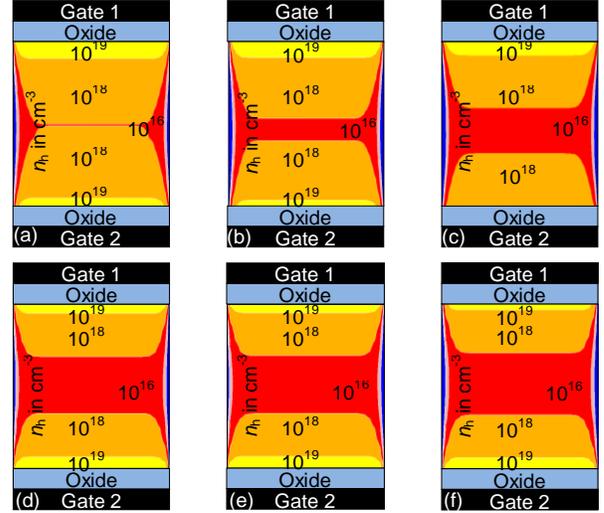

Fig. 4. Contour plots of hole concentration ($n_h$) during hold "1" condition with conventional architecture for hold time of (a) 10 ns, (b) 500 ms and (c) 1.5 s (*RT*) and with proposed architecture for hold time of (d) 10 ns (e) 500 ms and (f) 6.5 s (*RT*) with a fixed device and bias parameters.

Read operation is performed with drift-diffusion mechanism by applying a positive bias 1.2 V at the Gate 1, the back gate (Gate 2) in accumulation of hole with bias of -0.1 V and with a lower drain bias of 0.1 V to prevent the impact ionization in the device. In order to retain the state "1" and "0" for a longer duration, the AT is completely turned off, which helps to reduce the leakage current and enhance the retention of state "1" and "0", and thus, enhance the retention characteristic of the proposed DRAM cell. It is clearly observed from Fig. 4, which shows the contour plots of hole concentration ($n_h$) during hold "1" condition with conventional AM transistor with different hold time (Fig. 4(a) for 10 ns, Fig. 4(b) for 500 ms and Fig.4(c) for 1.5 s (*RT* of AM)) and proposed architecture (Fig. 4(d) for 10 ns, Fig. 4(e) for 500 ms and Fig.4(f) for 6.5 s (*RT* of proposed architecture)). The incorporation of AT in series with ME increases the *RT* and *SM* as AT is turned off during hold operation, which makes the ME turned off, and thus, reduces the recombination of holes.

The degradation/recombination of holes concentration with hold time in proposed architecture is lower as compared to conventional topology, which is beneficial for improvement in DRAM metrics such as *SM* and *RT*.

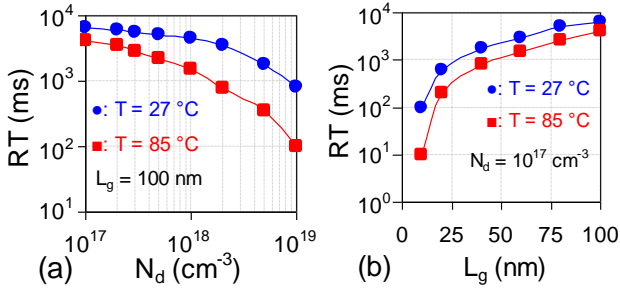

Fig. 5. (a) Variation in *RT* with $N_d$ at different temperature of 27 °C and 85 °C with gate length of 100 nm. (b) Variation in *RT* with gate length of memory element at different temperature at 27 °C and 85 °C with channel doping of $10^{17}$ cm$^{-3}$.

Capacitorless DRAM metrics are modulated through potential depth and carrier lifetime, and these are governed through doping of the channel [7]. Fig. 5(a) shows the variation of *RT* with channel doping ($N_d$) of the memory element (ME) with gate length of 100 nm at 27 °C and 85 °C. Although the potential depth and carrier lifetime of conventional AM/JL and proposed concept are same even though the proposed architecture archives higher retention time compared to conventional AM/JL device due to incorporation of access transistor in series. The maximum *RT* achieved are ~6.5 s and ~4.2 s at 27 °C and 85 °C, respectively, for $N_d$ of $10^{17}$ cm$^{-3}$ and reduces to ~800 ms and ~100 ms at 27 °C and 85 °C, respectively for $N_d$ of $10^{19}$ cm$^{-3}$ with gate length of 100 nm.

Downscaling of gate length degrades the DRAM cell performance due to short channel effects, BTBT, and reduction in storage region [9], [15]. Short channel effects can be controlled with accumulation mode transistor due to longer effective channel length, which shows better performance for DRAM cell [7]. Fig. 5(b) shows the variation in *RT* with gate length of ME for different temperatures of 27 °C and 85 °C with fixed gate length ($L_g$ = 100 nm) of AT and doping of $10^{17}$ cm$^{-3}$. The maximum *RT* achieved with $L_g$ of 100 nm are ~6.5 s and ~4.2 s at temperature of 27 °C and 85 °C, respectively, for $N_d$ of $10^{17}$ cm$^{-3}$. The proposed architecture shows better gate length scalability with *RT* of ~10 ms at a shorter $L_g$ (= 10 nm) for $N_d$ of $10^{17}$ cm$^{-3}$ at 85 °C.

Fig. 6 shows the transient analysis of proposed architecture to evaluate the energy consumption during write, hold and read operations. Table II summarizes the energy consumption for different DRAM operations. The energy required to perform write "1" is ~2.17×10$^{-3}$ pJ and for "0" it is almost negligible due to utilization of BTBT and forward bias mechanism phenomenon during operations, respectively. The read operations are performed with lower drain bias, therefore, it requires less energy. However, Read "1" operation consumes slightly more energy due to accumulation of higher carrier concentration in contrast to Read "0" in the potential well, which results in to higher current, and thus, consumes more energy. Also, hold operation consumes almost negligible energy because the transistor is turned off. Fig. 7 shows the comparison of our work with polished devices [8], [9], [18]–[20]. Filled symbols indicate 85 °C while empty symbols reflect 27 °C. Proposed architecture achieves higher *RT* compared to vertically stacked *n*-oxide-*p* transistor (*RT* = 600 ms [8], core-shell (CS) (*RT* = 5 ms (with same doping) [9], Silicon-with-partially-Insulating-layer on Silicon-on-Insulator (SISOI [18]) (*RT* = ~1.6 s) at $L_g$ = 100 nm due to lower leakage in the device. In comparison with GaAs based DRAM [19], the proposed DRAM achieves ~26 times higher *RT* with same doping. On the other hand, with thinner and separated transistor, proposed topology achieves ~6.6 times higher *RT* compared to [20] with doping of $10^{17}$ cm$^{-3}$ at 85 °C.

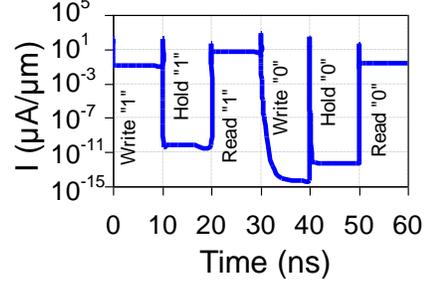

Fig. 6. Transient analysis of proposed capacitorless DRAM to evaluate the energy consumption during different operations.

TABLE II
ENERGIES REQUIRED FOR PROPOSED DRAM CELL

| Operation | $V_D$ (V) | Time (ns) | Energy (pJ) $E = I \times V \times T$ |
|---|---|---|---|
| Write "1" | 1.5 | 10 | 2.17×10$^{-3}$ |
| Hold | 0.0 | --- | 0.0 |
| Read | 0.1 | 10 | 7.35×10$^{-2}$ (Read "1") 4.01×10$^{-2}$ (Read "0") |
| Write "0" | 0.0 | 10 | 0.0 |

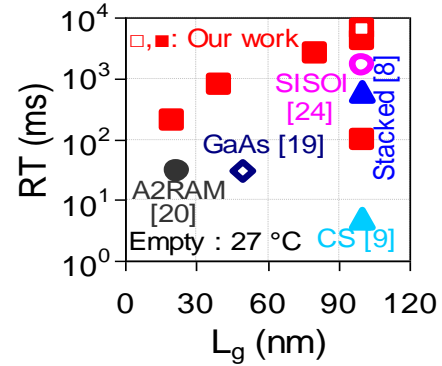

Fig. 7. Comparison of *RT* with published devices with gate length. Filled symbols indicate 85 °C while empty symbols reflect 27 °C.

## IV. CONCLUSION

This paper shows an opportunity to enhance the DRAM metrics of capacitorless DRAM cell by utilizing an access transistor in series with memory element (DRAM operations are performed with ME). By utilizing an access transistor reduces the generation and recombination holes during Hold operation compared to conventional architecture, and thus, enhances the retention characteristic and sense margin of the memory. Results show a better gate length scalability and achieves *RT* of 10 ms with gate length of 10 nm and doping of $10^{19}$ cm$^{-3}$ (JL) at 85 °C. The proposed architecture demonstrates better performance compared to the recently published work on capacitorless DRAM cell.